\documentclass[aps,prb,twocolumn,showpacs,groupedaddress,amsmath,amssymb]{revtex4}
\usepackage[dvips,usenames]{color}

\usepackage{graphicx}
\usepackage{hyperref}

\begin{document}

\input epsf

\title{Influence of nontrivial backgrounds and decoherence on vacuum decay}
\author{Friedemann Queisser}
\affiliation{Universit\"at zu K\"oln, Institut f\"ur Theoretische Physik, Z\"ulpicher Str. 77, D-50937 K\"oln, Germany}
\date{April 16, 2010}

\begin{abstract}
We present new solutions of vacuum decay in non-Minkowski spacetimes. 
On a de Sitter background, the tunneling between vacuum states with a nonvanishing energy difference has 
been considered, whereas in the special case of degenerate vacua we extended the treatment to a
power-law FRW-universe.
Additionally, we dicuss the vacuum decay in static spacetimes.
In the case of Minkowski spacetime, we include interaction with an
environmental field and analyze the effects of decoherence and the quantum-to-classical
transition of the nucleating bubble.
\end{abstract}

\maketitle

\section{Introduction}

The decay of metastable vacuum states has been of great interest since several decades.
Since the literature addressing this topic is enormous, we want to mention only some of the important results that have been found.

In the context of field theory, vacuum decay was first described using a semiclassical approach in \cite{COL77,Vol75,COLCAL77};
later on, gravitational effects on and of vacuum decay have been studied in \cite{COL80,Blau86,Ber87}.
The influence of finite temperature on vacuum decay has been addressed in \cite{Lin83}.

Based on quantum tunneling, important cosmological models were proposed, for example eternal inflation \cite{Lin82,VIL83},
the Hartle-Hawking-instanton \cite{HARHAW83,Vil82} or the Hawking-Moss instanton\cite{HawMoss82}.
The quantum creation of topological defects, e.g. strings and branes in a fixed space-time, have been discussed in \cite{GUTH91,GAR94}.

In the last years, several authors have suggested that string theory in four dimensions might have as many
as $10^{500}$ different vacua \cite{SUSSKIND03,DOUGLAS07}.
All these vacua considered in string theory are local minima of a very complicated potential, resulting from the huge amount of possible compactifications of the ten-dimensional, respectively eleven-dimensional theory, to four dimensions.
In this context the tunneling between different local minima is of great importance.

Many of the findings concerning quantum tunneling are based on the high symmetries of Minkowski and de Sitter space-time.
Coleman considered tunneling of scalar fields in a Minkowski background \cite{COL77} and in a 
closed de Sitter universe \cite{COL80}. 
In both cases, the $O(4)$-symmetry after Wick rotation has been used explicitely.

Here we want to generalize results of vacuum decay in curved backgrounds to various $O(3)$-symmetric
settings.
Although the chosen background is determined by Einstein's equation, we neglect the
backreaction of the (time-dependent) vacuum energy distribution on the curvature scalar.

The decay of metastable vacua is usually treated in the instanton picture which is of great success \cite{COLSYM}.
One assumes that the scalar field is initially located in a false vacuum $\phi_f$ and tunnels through 
a barrier into the true vacuum $\phi_t$.
In order to make the problem tractable, the field theoretical problem is reduced to a quantum mechanical problem by means of
symmetry considerations.
Both vacua denote classical minima of the potential and have to be distinguished
from the quantum mechanical ground state, which is given by the
symmetric superposition of $\phi_f$ and $\phi_t$.
One may ask under what circumstances it is allowed to ignore quantum mechanical superpositions
by considering only localized, i.e. ``classical'', vacuum states.
Analogous to this situation would be the localization of chiral molecules in left-handed or right-handed states,
although under certain circumstances, a superposition between both can be observed \cite{SCHLOSS,JZKGKS03}.
The localization of quantum states can only be justified with the influence of a system-environment
interaction; therefore one aim of this paper will be to investigate the influence of decoherence
on vacuum tunneling.
In the context of tunneling in the cosmic landscape, the influence of decoherence
 will be studied in a forthcoming publication \cite{Staro}.

This paper is organized as follows. 
In Section \ref{sec2} we summarize known results of vacuum decay.
We then present in Section \ref{sec3} some exact solutions of tunneling in fixed backgrounds.
The influence of decoherence on tunneling and the decay rate will be discussed in section \ref{sec4}.

\section{Vacuum decay in Minkowski space}\label{sec2}

The starting point is a scalar field theory 
\begin{eqnarray}\label{1}
S_\phi=\int d^4x\left(\frac{1}{2}\partial_\mu \phi\partial^\mu\phi -V(\phi)\right)\,,
\end{eqnarray}
where the potential $V(\phi)$ is assumed to have two local minima.
The difference between the energy densities in the localized vacua is denoted by $\epsilon$.

The Euclidean version of Feynman's path integral describing the transition
from $\phi_f$ to $\phi_t$ reads \cite{COLSYM}
\begin{eqnarray}\label{euclid}
\langle\phi_t|e^{-H T}|\phi_f\rangle=N\int D\phi e^{-S_\phi}\,,
\end{eqnarray}
where $T=it$ denotes the euclidean time, $H$ is the Hamiltonian corresponding to the action (\ref{1})
and $N$ is the normalization of the path integral.
The right hand side of equation (\ref{euclid}) can be evaluated in the semiclassical limit.
It follows, that the tunneling rate at lowest order is given by $\Gamma=A\exp(-2\Im(S))$, where $S$ denotes the classical tunneling action and $A$ includes the one-loop corrections \cite{COL77}.

Initially the field is constant in space, adopting the value $\phi_f$.
This situation is quantum mechanically unstable since the field can tunnel
through the barrier.
Due to the nucleation process, spatial regions with the field value $\phi_t$ are created spontaneously within
the initial configuration.

Assuming $O(4)$-invariance of the tunneling solution, 
the field theoretical problem reduces to a quantum mechanical problem with a single degree of freedom $\phi$ depending
on the four-dimensional radius $\rho$.
The effective particle moves in a potential $-V(\phi)$ from the false vacuum $\phi_f$ to the true vacuum $\phi_t$.

Following Coleman \cite{COL77}, the simplified euclidean action following from (\ref{1}) reads 
\begin{eqnarray}\label{2}
\Im(S)=-\frac{\pi^2}{4}\rho^4 \epsilon+\pi^2\rho^3S_1\,.
\end{eqnarray}
The first term in (\ref{2}) is a volume term originating from the field $\phi$ staying near $\phi_f$ until a very large ``time'' $\rho$.
The second term arises from the transition of the particle from $\phi_f$ to $\phi_t$ around the ``time'' $\rho$ where 
the soliton action $S_1$ depends on the concrete shape of the potential. 
In order to end up with a finite tunneling action, we demand that $V(\phi_f)=0$.
If $V(\phi_f)\neq0$, we consider $S_\phi-S_{\phi_f}$.

The action (\ref{2}) is minimized for $\rho=R_0=3S_1/\epsilon$, which leads to the famous result \cite{COL77}
\begin{eqnarray}\label{3}
\Im(S)=\frac{27\pi^2S_1^4}{4\epsilon^3}\,.
\end{eqnarray}
This specific tunneling action is valid in the so-called thin-wall approximation, meaning that the transition time between the vacua is small compared to the nucleation radius $R_0$.
Analytic continuation to Minkowski time leads to the conclusion that
the true vacuum bubble will expand almost instantly at the speed of light, since
\begin{eqnarray}\label{sol}
R(t)=\sqrt{R_0^2+t^2}\,.
\end{eqnarray}
It is possible to derive this result without referring explicitely to the $O(4)$-invariance of the 
problem after Wick rotation.
Assuming spherical symmetry of the expanding vacuum bubble, the action consists of a volume term involving
the difference between false and true vacuum and an integral over the two-dimensional surface of the sphere.
Using Minkowski time, the action reads \cite{KRAUSS96}
\begin{eqnarray}\label{4}
S_R&=&\int dt\left(\int_{|\mathbf{x}|\leq R} d^3x\sqrt{-\eta}\epsilon-\int_{|\mathbf{x}|=R} d^2x\sqrt{-\gamma}\sigma\right)\nonumber\\
&=&\int dt\left(\frac{4\pi R^3\epsilon}{3}-4\pi\sigma R^2\sqrt{1-\dot{R}^2}\right)\,,
\end{eqnarray}
where $\eta$ is the determinant of the Minkowski-metric and $\gamma$ is the determinant
of the induced metric on the surface of the sphere.
The relative minus sign in the action (\ref{4}) is due to the energy conservation
of the system: The difference of the energies in the nucleating region has to be balanced
by the negative energy of the surface of the sphere.
This allows the interpretation of $\sigma$ as surface tension.
The solution of the classical equations of motion coincides with (\ref{sol})
after the substitution $\sigma\rightarrow S_1$.

In contrast to (\ref{2}), the action (\ref{4}) can be generalized in a straightforward manner to problems without $O(4)$-symmetry.

\section{Tunneling in nontrivial backgrounds}\label{sec3}

The geometry of space-time is determined by Einstein's equations involving
the Ricci tensor and the energy momentum tensor.
Although any change of the matter distribution will have an impact on the
geometry, we will ignore this backreaction and consider the background
to be fixed.
Therefore we will discard any changes in the Einstein-Hilbert action due to the tunneling process.
The effective action determining the dynamics of the scalar field is a straightforward 
generalization of (\ref{4}) and reads (see also \cite{GUTH91})
\begin{eqnarray}\label{action}
S_R=\int dt\left(\int_{|\mathbf{x}|\leq R} d^3x\sqrt{-g}\epsilon-\int_{|\mathbf{\xi}|=R} d^2\xi\sqrt{-\gamma}\sigma\right)\,.
\end{eqnarray}
Here we have denoted the determinant of an arbitrary metric $g_{\mu\nu}$  by $g$, 
and $\gamma$ is the determinant of the induced metric
\begin{eqnarray}
\gamma_{ab}=g_{\mu\nu}\frac{\partial x^\mu}{\partial\xi^a}\frac{\partial x^\nu}{\partial\xi^b}\,,
\end{eqnarray}
where the $x^\mu(\xi)$ parametrize the space-time manifold on the sphere using
two coordinates $\xi^a$.

In the case of a Friedmann-Robertson-Walker universe with the line element
\begin{eqnarray}\label{5}
ds^2=a^2(y) (dy^2-dx^2-f^2(x) d\Omega^2)\,,
\end{eqnarray}
the action (\ref{action}) adopts the form
\begin{eqnarray}\label{6}
S_{x,FRW}&=&\int {dy}\bigg(4\pi\epsilon a^4(y)\int_0^{x(y)} dx'f^2(x')\nonumber\\
& & -4\pi\sigma a^3(y) f^2(x)\sqrt{1-\dot{x}^2(y)}\bigg)\,.
\end{eqnarray}
The conformal time is denoted with $y$ and the 
function $f$ is given by $x,\sin(x)$ and $\sinh(x)$ for flat, closed and open universes, respectively.
The coordinate of the bubble is given by the dimensionless function $x(y)$ and $\dot{x}$ denotes the 
derivative with respect to $y$.

For a given $a(y)$ it seems hopeless to find an analytic solution to the highly nonlinear 
equation of motion for $x(y)$.
Therefore we will solve the inverse problem: given a radius function $x(y)$, we
obtain solutions for a scale factor $a(y)$, among them solutions for physically reasonable cases.

The equation of motion resulting from (\ref{6}) reads
\begin{eqnarray}\label{7}
\frac{d}{dy}\left(\sigma a^3x^2\frac{\dot{x}}{\sqrt{1-\dot{x}^2}}\right)=\epsilon a^4x^2-2\sigma a^3x\sqrt{1-\dot{x}^2}\,.
\end{eqnarray}
In order to find solutions for the differential equation (\ref{7}), we assume the relation 
\begin{eqnarray}\label{8}
\sqrt{1-\dot{x}^2}=g(y)\dot{x}\,,
\end{eqnarray}
where $g(y)$ is chosen such that $g(y)\dot{x}$ is positive but otherwise arbitrary.
From here it follows that
\begin{eqnarray}\label{9}
\frac{\dot{a}}{a}-\frac{\dot{g}}{3g}-\frac{g a}{R_0}+\frac{2}{3}\frac{\partial_xf(x)}{f(x)\dot{x}}=0\,.
\end{eqnarray}
Equation (\ref{9}) has the general solution
\begin{eqnarray}\label{10}
a(y)=\frac{\left(\frac{g(y)}{g(y_0)}\right)^{1/3}e^{-F(y)}}
{C-\frac{1}{R_0}\int_{y_0}^y dy' \left(\frac{g(y')}{g(y_0)}\right)^{1/3}g(y')
e^{-F(y')}}\,,
\end{eqnarray}
with 
\begin{eqnarray}\label{11}
F(y)=\frac{2}{3}\int_{y_0}^y dy'\frac{\partial_xf(x)}{f(x)\dot{x}}\,.
\end{eqnarray}
The radius function reads
\begin{eqnarray}\label{12}
x(y)=\int_{\tilde{y}_0}^y dy'\frac{1}{\sqrt{1+g^2(y')}}\,.
\end{eqnarray}
Although the solution for arbitrary functions $x$ is thereby given in principle, the scale factor $a$, given by (\ref{10}), will have an awkward form in general.
The problem now is to find suitable functions $g$ and $x$ in order to obtain reasonable scale factors $a(y)$.

\subsection{De Sitter space}

De Sitter space is of great importance for the understanding of the early universe and maybe also for the 
future, since cosmological data suggest that our universe is dominated by dark energy with an
equation of state close to a cosmological constant \cite{Hinshaw08}.

De Sitter space is defined as a four-dimensional hyperboloid,
\begin{eqnarray}\label{13}
X_0^2-X_1^2-X_2^2-X_3^2-X_4^2=H^{-2}\,,
\end{eqnarray}
where $H$ denotes the Hubble parameter and $X_i$ are the coodinates in an auxiliary five-dimensional space.
It is possible to choose a flat, closed or open spatial slicing of the de Sitter space
leading to three different choices of coordinates.
In order to distinguish the conformal times of the different coordinate patches, we
denote them with $z$, $y$ and $w$ in case of the flat, closed and open spatial slicings, respectively.

The flat spatial sections of de Sitter space are defined as \cite{leshouches}
\begin{eqnarray}\label{14}
X_0&=&\frac{1}{2H}\left(-\frac{1}{z}+z-\frac{x^2}{z}\right)\,,\\
X_1&=&\frac{1}{2H}\left(-\frac{1}{z}-z+\frac{x^2}{z}\right)\,,\\
X_2&=&-\frac{x_1}{Hz}\,,\\
X_3&=&-\frac{x_2}{Hz}\,,\\
X_4&=&-\frac{x_3}{Hz}\,,
\end{eqnarray}
with $x_1^2+x_2^2+x_3^2=x^2$ and the conformal time $z$ running from $-\infty$ to $0$.
The line element reads
\begin{eqnarray}\label{15}
ds^2=a^2(z)(dz^2-dx^2-x^2d\Omega^2) 
\end{eqnarray}
with
\begin{eqnarray}\label{16}
a(z)=-\frac{1}{Hz}\,.
\end{eqnarray}
Using the equations (\ref{10}) and (\ref{12}) and the ansatz $g=\alpha/(z+\mathrm{constant})$, we find
\begin{eqnarray}\label{17}
x(z)=\sqrt{\alpha^2+\left(z+\frac{\alpha}{R_0 H}\right)^2}\,,
\end{eqnarray}
where the integration constant $\alpha$ is greater than zero.

In order to shrink the radius function (\ref{17}) to zero,
one has to continue $z$ analytically to the complex plane,
i. e. $z=-\alpha/(R_0 H)+iT$, with $T$ running from $\alpha$
to 0.
In order to determine the tunneling rate $\Gamma$, the action (\ref{6}) has to be evaluated
along the trajectory of $x$ given by the analytic continuation.
We find the expression
\begin{eqnarray}\label{18}
\Im(S)=\frac{\pi^2\epsilon}{3 H^4}\frac{\left(1-\sqrt{1+R_0^2H^2}\right)^2}{\sqrt{1+R_0^2H^2}}\,,
\end{eqnarray}
which is independent of $\alpha$
and coincides with the result already found by Simon \textit{et al.}\cite{ADAMEK09}.
These authors found the tunneling rate (\ref{18}) by means of a differential equation obtained from
(\ref{7}) after some simplifying approximations, whereas here we have shown that the tunneling
amplitude can be obtained directly from (\ref{7}) and without approximations.
In the limit $H\rightarrow 0$, equation (\ref{18}) coincides with (\ref{3}).
In the limit $\epsilon\rightarrow 0$, one obtains $\Im(S)=\pi^2\sigma/H^3$, which is
the result for the nucleation of a domain wall separating two degenerate vacua found
by Basu \textit{et al.}\cite{GUTH91}.

The physical bubble radius is
\begin{eqnarray}\label{19}
R_\mathrm{phys}=a x=-\frac{\sqrt{\alpha^2+\left(z+\frac{\alpha}{R_0 H}\right)^2}}{Hz}\,,
\end{eqnarray}
and the radius at nucleation, $R_\mathrm{nucl,flat}=R_0$, is independent of $\alpha$.
Since the action is invariant under the rescaling $z\rightarrow \alpha z$ and $x\rightarrow \alpha x$,
it is possible to eliminate $\alpha$.
This explains why (\ref{18}) does not depend on this parameter.

\begin{figure}
\includegraphics{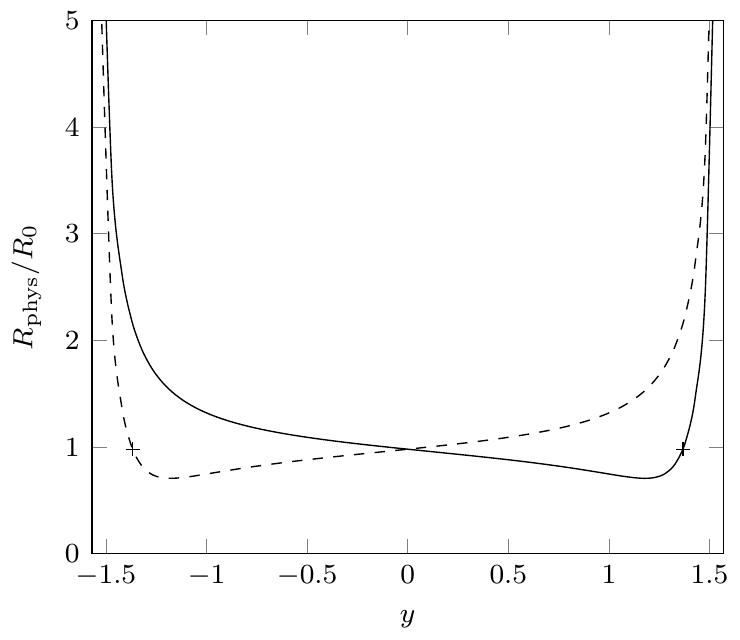}
\caption{Choosing a slicing of de Sitter space with positive curvature, we find for the same values
of $R_0 H$ and $\alpha$ two different solutions that can be obtained from each other by reflection
with respect to the $R_\mathrm{closed}/R_0$-axis.
The marks on the curves label the nucleation radius and the nucleation time defined by $R_\mathrm{phys}(y_\mathrm{nucl})=R_\mathrm{nucl,closed} $. 
}\label{fig1}
\end{figure}

Now we may transform this result to the closed spatial sections of de Sitter space-time, which
are parametrized by \cite{leshouches}
\begin{eqnarray}\label{20}
X_0&=&\frac{\sin(y)}{H\cos(y)}\,,\\
X_1&=&\frac{1}{H\cos(y)}\cos(\chi)\,,\\
X_2&=&\frac{1}{H\cos(y)}\sin(\chi)\cos(\Theta)\,,\\
X_3&=&\frac{1}{H\cos(y)}\sin(\chi)\sin(\Theta)\cos(\phi)\,,\\
X_4&=&\frac{1}{H\cos(y)}\sin(\chi)\sin(\Theta)\sin(\phi)\,.
\end{eqnarray}
The line element reads
\begin{eqnarray}\label{21}
ds^2=a^2(y)(dy^2-d\chi^2-\sin^2(\chi)d\Omega^2) 
\end{eqnarray}
with
\begin{eqnarray}\label{22}
a(y)=\frac{1}{H\cos(y)},\quad y\in\left(-\frac{\pi}{2},\frac{\pi}{2}\right)\,.
\end{eqnarray}
Transforming the solution (\ref{17}) to a closed de Sitter universe and 
using (\ref{10}), we obtain the physical radius
\begin{eqnarray}\label{23}
R_\mathrm{phys}&=&a\sin(\chi)\nonumber\\
&=&\hspace{-.1cm}\frac{1}{H\cos(y)}\sqrt{1-\left(\frac{1-A^2}{1+A^2}\right)^2\sin^2(y-y_0)}
\end{eqnarray}
with 
\begin{eqnarray}\label{24}
A=\alpha\sqrt{1+\frac{1}{R_0^2 H^2}},\quad\alpha>0,
\end{eqnarray}
and 
\begin{eqnarray}\label{25}
y_0=\pm\arcsin\left(\frac{2 }{R_0 H}\frac{A}{1-A^2}\right)\,.
\end{eqnarray}
The solution with positive $y_0$ has a minimum in the contracting branch
of the closed de Sitter universe at
\begin{eqnarray}\label{26}
y_\mathrm{min}=-\arccos\frac{\left(1+\frac{1}{R_0^2 H^2}\right)^2A}{\sqrt{A^2+\frac{1}{4R_0^2 H^2}(1+6A^2+A^4)}}
\end{eqnarray}
with the minimal radius
\begin{eqnarray}\label{27}
R_\mathrm{min}=\frac{R_0}{\sqrt{1+R_0^2 H^2}}\,.
\end{eqnarray}
The solution with negative $y_0$ has the minimum value $R_\mathrm{min}$ in the expanding branch at $-y_\mathrm{min}$.
In contrast to the flat de Sitter solution, the nucleation radius depends on $H$ and $\alpha$.
We have
\begin{eqnarray}\label{28}
R_\mathrm{nucl,closed}=R_0\frac{1-A^2}{1+A^2}
\end{eqnarray}
at the nucleation time
\begin{eqnarray}\label{29}
y_\mathrm{nucl}=\arccos\left(\frac{2}{R_0 H}\frac{A}{1-A^2}\right)
\end{eqnarray}
for the solution with the minimum in the expanding branch.
The corresponding solution with the minimum in the contracting
branch adopts the value (\ref{28}) at the nucleation time $-y_\mathrm{nucl}$.
For this solution, it is possible that the physical nucleation radius is \textit{larger} than the physical radius
at subsequent times (see Fig.\ref{fig1}).
The nucleation radius is determined by the requirement 
that the \textit{comoving} radius adopts a minimum value for some real nucleation time.
Subsequently, one choses the analytical continuation to complex time such that comoving bubble radius shrinks to zero.

The value of $A$ is constrained by the condition that $y_0$ given by (\ref{25}) has to be real. 
From this condition and equation (\ref{27}), we conclude that the
nucleation radius is constrained by the relation
\begin{eqnarray}\label{30}
R_\mathrm{min}\leq R_\mathrm{nucl,closed}\leq R_\mathrm{nucl,flat}\,.
\end{eqnarray}
In order to shrink the bubble to zero we use complex time $y=y_\mathrm{nucl}+iU$, with $U$ running from $\mathrm{arcosh}\frac{1+A^2}{1-A^2}$ to 0.

A third possibility is the spatially open slizing of de Sitter space using the coordinates \cite{leshouches}
\begin{eqnarray}\label{31}
X_0&=&-\frac{\cosh(\psi)}{H\sinh(w)}\,,\\
X_1&=&-\frac{\cosh(w)}{H\sinh(w)}\,,\\
X_2&=&-\frac{1}{H\sinh(w)}\sinh(\psi)\cos(\Theta_2)\,,\\
X_3&=&-\frac{1}{H\sinh(w)}\sinh(\psi)\sin(\Theta_2)\cos(\Theta_3)\,,\\
X_4&=&-\frac{1}{H\sinh(w)}\sinh(\psi)\sin(\Theta_2)\sin(\Theta_3)\,,
\end{eqnarray}
and the scale factor
\begin{eqnarray}\label{32}
a(w)=-\frac{1}{H\sinh(w)},\quad w\in(-\infty,0)\,.
\end{eqnarray}
Transforming into these coordinates and using (\ref{10}), we find
\begin{eqnarray}\label{33}
\hspace{-.3cm}R_\mathrm{phys}&=&a \sinh(\psi)\nonumber\\
& &\hspace{-1.2cm}=-\frac{1}{H\sinh(w)}\sqrt{\left(\frac{1+A^2}{1-A^2}\right)^2\cosh^2(w+w_0)-1}
\end{eqnarray}
with
\begin{eqnarray}\label{34}
w_0=\mathrm{arsinh}\left(\frac{2}{R_0 H}\frac{A}{1+A^2}\right)>0\,.
\end{eqnarray}
The bubble is nucleating at the radius
\begin{eqnarray}\label{35}
R_\mathrm{nucl,open}=R_0\frac{1+A^2}{1-A^2}
\end{eqnarray}
and the time
\begin{eqnarray}\label{nuclopen}
w_{\mathrm{nucl,open}}=-w_0\,.
\end{eqnarray}
Obviously, the size of the nucleation radius is constrained by
\begin{eqnarray}
R_\mathrm{nucl,open}\geq R_\mathrm{nucl,flat}\,.
\end{eqnarray}
The analytic continuation of the time is $w=w_\mathrm{nucl}+iV$, and $V$ runs from $\arccos\frac{1-A^2}{1+A^2}$ to 0.

It is remarkable that the tunneling rate for the flat, closed and open slicing is determined by the same expression (\ref{18}).
However, this does not mean that the physical size of the nucleating bubble
is the same in the three different cases, as can be seen from equations (\ref{28}) and (\ref{35}).
For the flat slicing of de Sitter space, the nucleation radius is independent of the nucleation time
due to the scale invariance of the action (\ref{6}).
In contrast, the nucleation radius for the closed and open slicing of de Sitter 
space depends explicitely on the parameter $\alpha$ respectively the 
nucleation times (\ref{29}) and (\ref{nuclopen}), which reflects
the loss of scale invariance in these cases.

\subsection{Power-Law expansion in a spatially flat universe}
It is also possible to find solutions for a power-law scale factor if one restricts
to the case $\epsilon=0$, i.e. two degenerate vacua.

In order to obtain the de Sitter universe as a limit for $n\rightarrow\infty$, we choose
the scale factor of the form
\begin{eqnarray}\label{36}
a(t)=\frac{1}{H}\left(1+\frac{H t}{n}\right)^n\,,
\end{eqnarray}
where $t$ is the cosmological time.
Changing the parametrization from $t$ to conformal time $z$, the scale factor reads
\begin{eqnarray}\label{37}
a(z)=\frac{1}{H}\left(-\frac{n-1}{n}z\right)^{-\frac{n}{n-1}}\,.
\end{eqnarray}
Unfortunately it is not possible to obtain an analytic expression for $x(z)$
such that the scale factor is exactly of the form given by equation (\ref{37}).
We will merely find a scale factor which coincides with the expression (\ref{37})
for small $z$, i.e. large cosmological times $t$.

In order to obtain a power law behaviour one has to choose for the radius function
of the bubble;
\begin{eqnarray}\label{38}
x(z)=\sqrt{\alpha^2+\frac{2n-2}{2n+1}z^2}\,,
\end{eqnarray}
which coincides for $\epsilon\rightarrow0$ and $n\rightarrow\infty$ with the result (\ref{17}).
The scale factor then reads
\begin{eqnarray}\label{39}
a(z)=\frac{1}{H}\hspace{-.1cm}\left(-\frac{n-1}{n}z\right)^{-\frac{n}{n-1}}\hspace{-.1cm}\left(1+\frac{6(n-1)}{(2n+1)^2}\frac{z^2}{\alpha^2}\right)^{1/6}\,,
\end{eqnarray}
which has the expression (\ref{37}) as limit for $z\rightarrow 0$.

Using this result, we can calculate the tunneling amplitude via the instanton action.
In order to increase the bubble radius from $0$ to $\alpha$, the conformal time has to run
from $i\alpha\sqrt{(2n+1)/(2n-2)}$ to $0$.

We find for the imaginary part of the action (\ref{6})
\begin{eqnarray}\label{40}
\Im(S)&=&\sigma\alpha^3 \pi^{3/2}\left(\frac{n-1}{n}\alpha H\right)^{-\frac{3n}{n-1}}
\left(\frac{2(n-1)}{2n+1}\right)^{\frac{4n-1}{2(n-1)}}\hspace{-.1cm}\times\nonumber\\
& &\hspace{-1.3cm}\times\frac{n-1}{n-4}\sin\left(\frac{\pi(2n+1)}{2(n-1)}\right)\frac{\Gamma\left(-\frac{1}{n-1}\right)}{\Gamma\left(\frac{1}{2}-\frac{1}{n-1}\right)}
\xrightarrow{n\rightarrow\infty} \frac{\pi^2\sigma}{H^3}\,.
\end{eqnarray}
For $n\rightarrow1$, the expression oscillates rapidly, which means that the WKB approximation breaks down.
This is due to the conformal-time parametrization of the scale factor.
Note that the tunneling rate depends here on the size $\alpha$ of the bubble; the scale invariance is established only 
for $n\rightarrow\infty$.

\subsection{Bubble expansion without tunneling}

The nucleation of a vacuum bubble has been described so far through 
an increase of the radius from from zero to $R_\mathrm{nucl}$
using analytical continuation of the time.
One may ask whether it is possible to find solutions of (\ref{7})
where no analytical continuation is necessary for the increase
of the vacuum bubble from zero.
This can be achieved by choosing the scale factor such that the tunneling
barrier vanishes.

\begin{figure}
\includegraphics{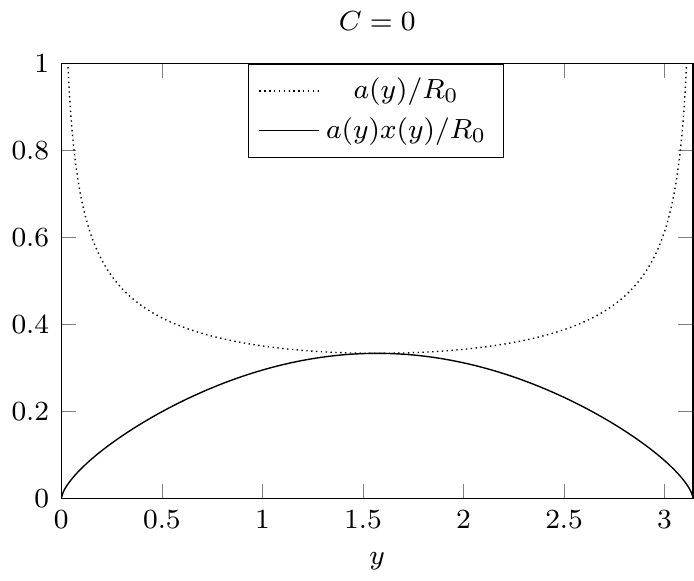}
\caption{The physical bubble radius and the scale factor are plotted as functions of the conformal time, 
the constant of integration was chosen to be $C=0$.
This solution allows only an increase of the radius from zero, if the scale factor is infinitely large at $y=0$.
}\label{fig10}
\end{figure}

We consider a flat FRW-universe and
\begin{eqnarray}
g(y)=\tan(y)\,,
\end{eqnarray}
giving the radius function for the bubble to be
\begin{eqnarray}
x(y)=\sin(y)\,.
\end{eqnarray}
Using equation (\ref{10}) we find
\begin{eqnarray}
a(y)=\frac{R_0 |\cot(y)|^{1/3}}{3|\cos(y)|^{1/3}
F_{21}\left(\frac{1}{6},\frac{1}{6},\frac{7}{6},\cos^2(y)\right)+C}\,,
\end{eqnarray}
where $F_{21}$ is a hypergeometric function, and $C$ is a constant greater than or equal to zero (see Fig.\ref{fig10}).
If the integration constant is chosen to be zero, the radius of the vacuum bubble  
increases from zero at $y=0$, given that the scale factor is infinitely large at $y=0$.

\begin{figure}
\includegraphics{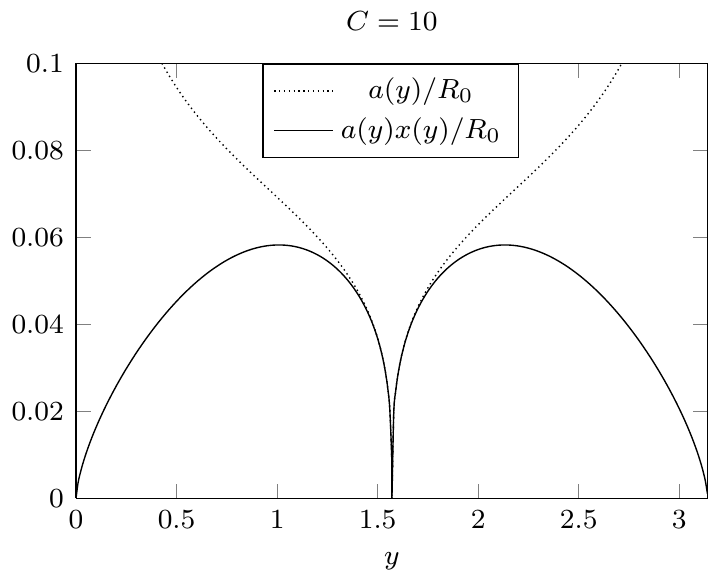}
\caption{The physical bubble radius and the scale factor are plotted as functions of the conformal time,
the constant of integration was chosen to be $C=10$.
The scale factor and the bubble radius increase from zero starting at $y=\pi/2$.}\label{fig11}
\end{figure}

Choosing any $C>0$, we find that the scale factor and the bubble radius
grow from zero starting at $y=\pi/2$.
Afterwards, the scale factor becomes infinitely large at $y=\pi$ whereas the bubble radius
decreases to zero at $y=\pi$ after adopting some maximum value (see fig.\ref{fig11}).
Due to the symmetry of the solution, there is also a branch where the bubble radius
starts at zero for an infinite scale factor and decreases to zero again at $y=\pi/2$.
Therefore we have obtained solutions that allow an expansion of a true vacuum bubble 
without a previous tunneling process.

\subsection{Static space-times}

Solutions for vacuum decay in an $O(3)$-symmetric background with an explicit timelike Killing symmetry
are easy to obtain if the center of the vacuum bubble coincides with the
fixed point of the rotation symmetry.

The line element of a static space-time has the form 
\begin{eqnarray}\label{41}
ds^2=f(r)dt^2-f^{-1}(r)dr^2-r^2 d\Omega^2\,,
\end{eqnarray}
where $f$ is some function which depends only on $r$.
From equation (\ref{action}) we find
\begin{eqnarray}\label{42}
S=\frac{4\pi}{3}\epsilon R_0^4\int dy\, x^2\left(x-\sqrt{f-f^{-1}\dot{x}^2}\right)\,,
\end{eqnarray}
where we have introduced the dimensionless radius function $x=R/R_0$ and the dimensionless
time $y=t/R_0$.
From conservation of energy which is due to the timelike Killing symmetry, we find the first-order differential equation
\begin{eqnarray}\label{43}
\dot{x}=f\sqrt{1-\frac{f}{x^2}}\,.
\end{eqnarray}
Whether it is possible to find a solution of the form $x=x(y)$ depends on the function $f$.
The imaginary part of the tunneling action is given by
\begin{eqnarray}\label{44}
\Im(S)=\frac{4\pi}{3}R_0^4\int_{x_1}^{x_2} dx x^2 f^{-1}\sqrt{f-x^2}\,,
\end{eqnarray}
where $x_1$ and $x_2$ are given by the two positive roots of $f-x^2$.
These roots are exactly the turning points of the tunneling trajectory
through an effective particle potential.
Since this cannot be seen from (\ref{42}) directly due to the non-standard
form of the action, we will switch to the Hamiltonian formalism.

In order to get rid of the square root in the action we parametrize
the action with an affine parameter $\lambda$, \cite{brink}
\begin{eqnarray}\label{45}
S&=&\frac{4\pi}{3}\epsilon R_0^4\int d\lambda\, x^2\Bigg(x\frac{dy}{d\lambda}\nonumber\\
& &\hspace{1cm}-\sqrt{f\left(\frac{dy}{d\lambda}\right)^2-f^{-1}\left(\frac{dx}{d\lambda}\right)^2}\Bigg)\,.
\end{eqnarray}
Introducing an auxiliary
variable $\nu$ leads to an action classically equivalent to (\ref{42}):
\begin{eqnarray}\label{46}
\tilde{S}&=&\frac{4\pi}{3}\epsilon R_0^4\int d\lambda x^2\bigg(x\frac{dy}{d\lambda}\nonumber\\
& &-\frac{f(dy/d\lambda)^2-f^{-1}(dx/d\lambda)^2}{2\nu}-\frac{\nu}{2}\bigg)\,.
\end{eqnarray}
The corresponding Hamiltonian constraint reads
\begin{eqnarray}\label{47}
H&=&\frac{3}{8\pi\epsilon R_0^4}\bigg(\frac{f P_x^2}{x^2}-\frac{1}{f x^2}\left(P_y-\frac{4\pi}{3}\epsilon R_0^4 x^3\right)^2+\nonumber\\
& &+\left(\frac{4\pi}{3}\epsilon R_0^4\right)^2 x^2\bigg)\approx 0\,.
\end{eqnarray}
This gives rise to the potential of an effective particle potential
\begin{eqnarray}\label{48}
V(x)=\frac{2\pi}{3}\epsilon R_0^4 x^2 f^{-1}(f-x^2)\,. 
\end{eqnarray}
Using this potential we are able to interpret the results easily.
Due to the timelike Killing vector field we have a conserved energy which
constrains the Hamiltonian to zero.
If the potential is greater than zero, the particle has to tunnel through the 
barrier.
If $V(x)<0$ for all $x$, the particle will leave the false vacua without tunneling.

\begin{figure}
\includegraphics{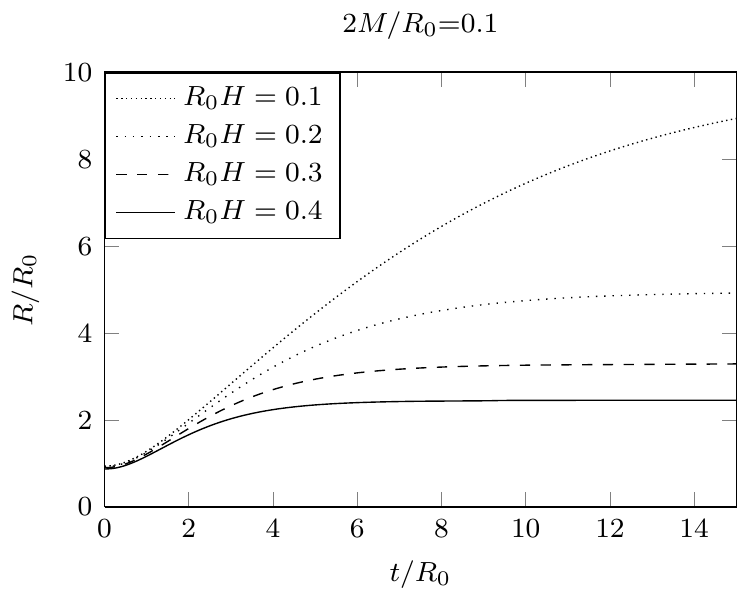}
\caption{Schwarzschild-de Sitter space-time: The vacuum bubbles nucleating at a radius larger than the inner horizon of the
Schwarzschild-de-Sitter-space-time reach the outer horizon in the limit of infinite times. 
}\label{fig2}
\end{figure}

\begin{figure}
\includegraphics{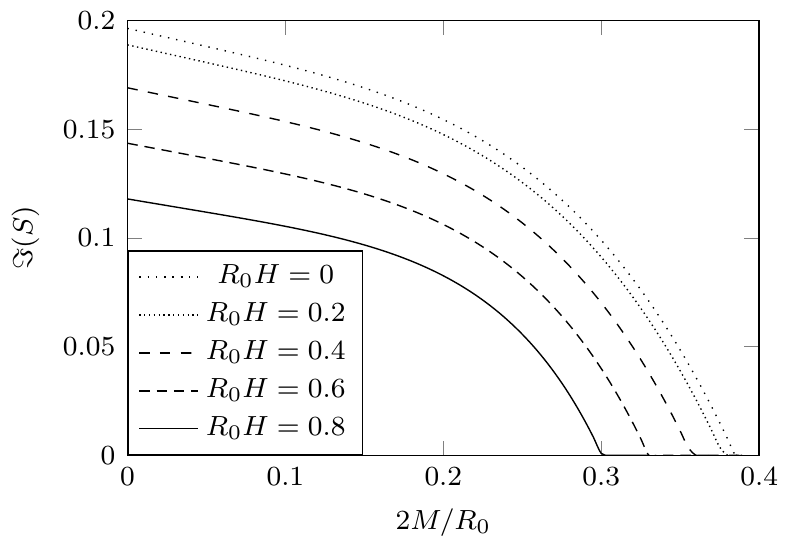}
\caption{Schwarzschild-de Sitter space-time: The imaginary part of the tunneling action decreases with growing black hole mass $M$.}\label{fig4}
\end{figure}

\begin{figure}
\includegraphics{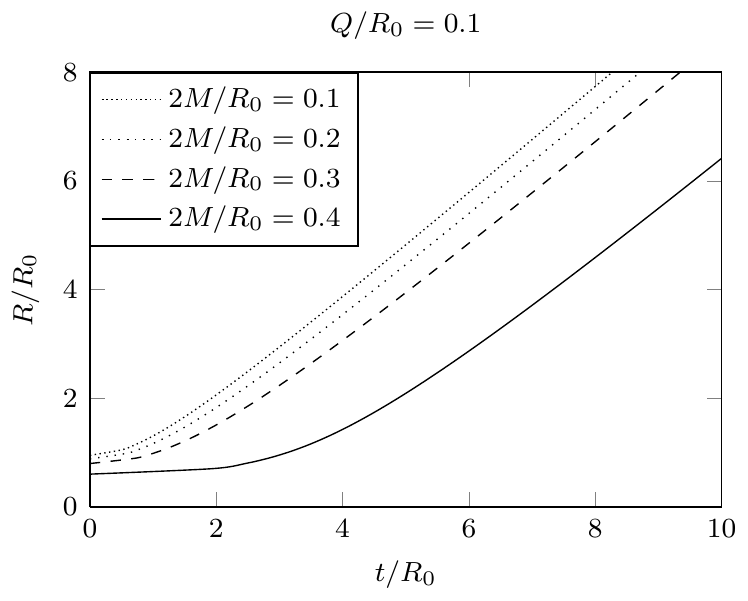}
\caption{Reissner-Nordstr\"om space-time: The expansion of the vacuum bubbles starting at a nucleation radius larger than the outer horizon is growing to infinity.}\label{fig3}
\end{figure}

\begin{figure}
\includegraphics{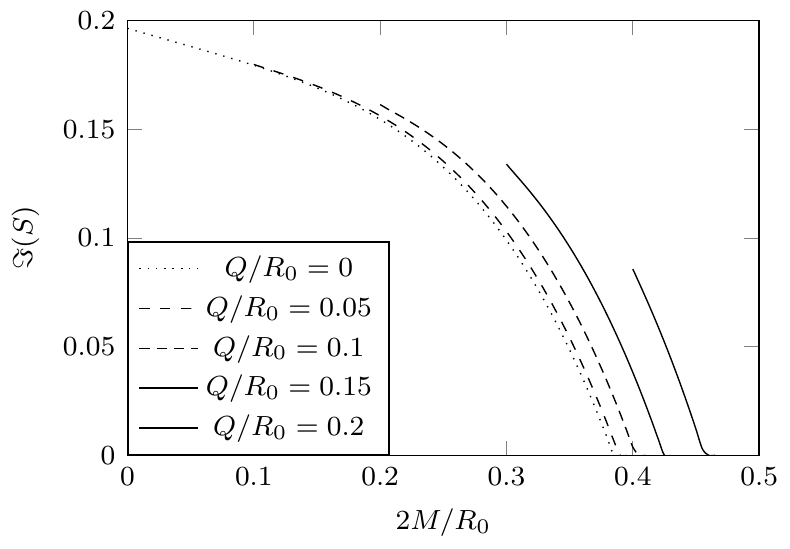}
\caption{Reissner-Nordstr\"om space-time: The imaginary part of the tunneling action decreases with growing black hole mass $M$.
Since we do not consider naked singularities, we discard the values for $\Im(S)$ if $|Q|>M$.}\label{fig5}
\end{figure}

We now consider two concrete cases: the Schwarzschild-de Sitter and the Reissner-Nordstr\"om space-times.
For the Schwarzschild-de Sitter space-time we have
\begin{eqnarray}\label{49}
f(x)=1-\frac{2 M}{R_0 x}-H^2R_0^2 x^2 \,.
\end{eqnarray}
From this follows that the particle has to tunnel between
\begin{eqnarray}\label{50}
x_1=\frac{2}{\sqrt{3(1+R_0^2 H^2)}}\cos\left(\frac{1}{3}\arccos\left(\beta\right)-\frac{2\pi}{3}\right)
\end{eqnarray}
and
\begin{eqnarray}\label{51}
x_2=\frac{2}{\sqrt{3(1+R_0^2 H^2)}}\cos\left(\frac{1}{3}\arccos\left(\beta\right)\right)
\end{eqnarray}
with
\begin{eqnarray}\label{52}
\beta=-\frac{\sqrt{27}M}{R_0}\sqrt{1+R_0^2 H^2}\,.
\end{eqnarray}
The tunneling always occurs between the two horizons of Schwarzschild-de Sitter space.
The trajectories of the bubble shell after tunneling are plotted in Fig.\ref{fig2} for different parameters.
We see that close to the outer horizon the velocity decreases to zero.

Furthermore, the imaginary part of the tunneling action decreases for increasing
black hole mass, since the barrier is lowered for increasing $M$.
In Fig.\ref{fig4} the imaginary part of the action is plotted for different values of $M$ and $H$.
The barrier vanishes completely for
\begin{eqnarray}\label{53}
M>\frac{R_0}{\sqrt{27(1+R_0^2 H^2)}}\,.
\end{eqnarray}
Our second example is the Reissner-Nordstr\"om-space-time defined by
\begin{eqnarray}\label{54}
f(x)=1-\frac{2 M}{R_0 x}+\frac{Q^2}{R_0^2 x^2}\,.
\end{eqnarray}
If $|Q|<M$ holds, the line element describes a black hole with charge,
if $|Q|>M$, the space-time describes a naked singularity. 
In the following we want to discard the latter case.

For $|Q|<M$, tunneling between $x_1$ and $x_2$ occurs only if $f-x^2$ has three positive roots.
If there is only a single positive root of $f-x^2$, we find that a tunneling solution exists only 
behind the inner horizon and is not visible for an observer outside.
Under the restrictions that $|Q|<M$ and that three positive roots exist, we find after some algebra
\begin{eqnarray}\label{55}
Q^2<M^2<\frac{R_0^2}{54}\left(1+36 \frac{Q^2}{R_0^2} +\left(1-12\frac{Q^2}{R_0^2}\right)^{3/2}\right)\,,
\end{eqnarray}
which implies $Q^2<R_0^2/16$.
The turning points of the potential are given by
\begin{eqnarray}\label{56}
x_{1/2}&=&\sqrt{\frac{1}{6}\left(1+\Re\left(\frac{\Delta}{2}\right)^{1/3}\right)}\\
& &\hspace{-1cm}\pm \sqrt{\frac{1}{3}-\frac{1}{6}\Re\left(\frac{\Delta}{2}\right)^{1/3}-
\frac{M}{R_0}\sqrt{\frac{3}{2}}\left(1+\Re\left(\frac{\Delta}{2}\right)^{1/3}\right)^{-1/2}}\nonumber
\end{eqnarray}
with
\begin{eqnarray}\label{57}
\Delta&=&-2-72\frac{Q^2}{R_0^2}+108\frac{M^2}{R_0^2}\\
& &+i\sqrt{4\left(1-12\frac{Q^2}{R_0^2}\right)^3-\left(2+72\frac{Q^2}{R_0^2}-108\frac{M^2}{R_0^2}\right)^2}\nonumber\,.
\end{eqnarray}
In Fig.\ref{fig3} we plot the classical trajectory of the bubble after tunneling. 
The imaginary part of the action for different values of $M$ and $Q$ is depicted in Fig.\ref{fig5}.

\section{Interaction with external degrees of freedom}\label{sec4}

\subsection{The system-environment interaction}
So far we have treated the field $\phi$ as an isolated system
and considered the tunneling from the localized state $\phi_f$
to the localized state $\phi_t$.
However, in general the field could be in a superposition of these
two localized states.
If this were the case, one could not justify the simple semiclassical
picture of the field moving along a trajectory through the barrier. 

Here we will show that the localization of the field 
in the potential wells can be understood through decoherence
due to the interaction of a field with external degrees of freedom
in the spirit of \cite{Staro}.

The environment will by modelled by a scalar field with the action
\begin{eqnarray}\label{58}
S_\mathrm{bath}=\frac{1}{2}\int dt\int d^3x\left(\partial_\mu\psi\partial^\mu\psi-m^2\psi^2\right)\,,
\end{eqnarray}
where the mass $m$ is a free parameter. 

For the interaction between the fluctuations $\psi$ and the bounce field $\phi$ we choose
\begin{eqnarray}\label{59}
S_\mathrm{int}=\int dt\int d^3x g\phi\psi\,.
\end{eqnarray}
This can be written as
\begin{eqnarray}\label{60}
S_\mathrm{int}&=&g\int d^4x \phi\psi\\
&=&g\int d^4x \phi_f\psi-g\int dt\int_{|\mathbf{x}|<R} d^3x(\phi_f-\phi_t)\psi\nonumber\,,
\end{eqnarray} 
where we have neglected the small transition region between $\phi_f$ and $\phi_t$.
After Fourier transforming the environmental scalar field, the interaction reads
\begin{eqnarray}\label{61}
S_\mathrm{int}\hspace{-0.1cm}&=&\hspace{-0.1cm}g(\phi_t-\phi_f)\hspace{-0.1cm}\int\hspace{-0.1cm} dt\int_{|\mathbf{x}|<R} d^3x \sum_\mathbf{k}e^{i\mathbf{kr}}\psi_\mathbf{k}\\
 \hspace{-0.1cm}&=& \hspace{-0.1cm}g(\phi_t-\phi_f)\hspace{-0.1cm}\int\hspace{-0.1cm} dt \sum_\mathbf{k}\frac{4\pi}{k^3}\left(\sin(k R)-k R\cos(k R)\right)\psi_\mathbf{k}\nonumber\,,
\end{eqnarray}
where we have neglected the constant first term in the second line of equation (\ref{60}).
The radius $R$ will be interpreted as a quantum variable like in the action (\ref{4}).
The interaction between $R$ and $\psi$ grows with $kR$, i.e. modes with short wavelengths are able to resolve
the vacuum bubble better than modes with longer wavelength, as expected.

\subsection{Effective two-state system}

An approximation for the tunneling between two local minima is a two-level system.
In order to estimate whether the instanton picture is justified we will use this simplification.
We approximate the system and the environment by the Hamiltonian
\begin{eqnarray}\label{62}
H_\mathrm{total}&=&\begin{pmatrix}
                  \frac{4\pi}{3}R_0^3V(\phi_f) & \Gamma\\
		  \Gamma & \frac{4\pi}{3}R_0^3V(\phi_t)
                 \end{pmatrix}\nonumber\\
& &+\frac{1}{2} \int d^3x \left(\Pi_\psi^2+(\nabla \psi)^2+m^2\psi^2\right)\nonumber\\
& &-g\begin{pmatrix}
                  \phi_f & 0\\
		  0& \phi_t
                 \end{pmatrix}\int_{|\mathbf{x}|<R_0} d^3 x\,\psi \,.
\end{eqnarray}
The first term in equation (\ref{62}) describes the transition between the states ``A bubble of radius $R_0$ has energy density $V(\phi_f)$''
and ``A bubble of radius $R_0$ has energy density $V(\phi_t)$''\,.
The second term of equation (\ref{62}) is the bath Hamiltonian, and the last term describes the environment measuring the 
system to be located at $\phi_f$ respectively $\phi_t$.

The master equation for the reduced density matrix in the Schr\"odinger picture reads in
the Redfield approximation \cite{SCHLOSS}
\begin{eqnarray}\label{63}
\dot{\rho}_S(t)&=&-i[H_0,\rho_S]\\
& &-\mathrm{tr}_B\int_0^t ds [H_\mathrm{int},[H_\mathrm{int}(s-t),\rho_S(t)\otimes\rho_B]]\nonumber
\end{eqnarray}
with
\begin{eqnarray}\label{64}
H_\mathrm{int}&=&-4\pi g\begin{pmatrix}
                  \phi_f & 0\\
		  0& \phi_t
                 \end{pmatrix}\sum_\mathbf{k}\int_0^{R_0}dr\frac{r}{k}\sin(kr)\psi_\mathbf{k}\nonumber\\
&\equiv&\sum_\mathbf{k}M_\mathbf{k}(R_0)\psi_\mathbf{k}\,.
\end{eqnarray}
The reduced density matrix $\rho_S$ contains all available information about the effective two-state system.

Ignoring the free dynamics of the density matrix, equation (\ref{63}) becomes
\begin{eqnarray}\label{65}
\dot{\rho}_S&=&-\int_0^t \hspace{-0.2cm}ds\sum_\mathbf{k}A_\mathrm{k}(t-s)[M_\mathrm{k}(R_0),[M_\mathrm{k}(R_0),\rho_S]]\\
& &+i\int_0^t \hspace{-0.2cm}ds \sum_\mathbf{k}B_\mathrm{k}(t-s)[M_\mathrm{k}(R_0),\{M_\mathrm{k}(R_0),\rho_S\}]\nonumber\,.
\end{eqnarray}
The functions $A_k$ and $B_k$ are defined by
\begin{eqnarray}\label{66}
\langle\hat{\psi}_\mathbf{k}\hat{\psi}_\mathbf{k}(s-t)\rangle&=&
\frac{1}{2\mathcal{V}k}[\cos(k(t-s))-i\sin(k(t-s))]\nonumber\\
&\equiv&A_k(t-s)-iB_k(t-s)\,,
\end{eqnarray}
where $\mathcal{V}$ denotes the quantization volume.
We have restricted ourselves here to vanishing temperature.
The second line of (\ref{65}) contains a contribution to the unitary dynamics and will be ignored from here.
Using this approximation we integrate equation (\ref{65}) and find that the off-diagonal elements of the density matrix decay according to
\begin{eqnarray}\label{67}
\rho_{S,01}(t)=\rho_{S,01}(0)e^{-4g^2(\phi_f-\phi_t)^2R_0^6 h(R_0,t)}
\end{eqnarray}
with
\begin{eqnarray}\label{68}
 h(R_0,t)&=&\hspace{-.1cm}\frac{13 t^2}{180R_0^2}+\frac{ t^4}{720R_0^4}
 +\ln t\left(\frac{ t^4}{48R_0^4}-\frac{t^6}{1440R_0^6}\right)\nonumber\\
&+ &\hspace{-.1cm}\ln\left|\frac{t+2 R_0}{t-2 R_0}\right|\left(\frac{ t}{15R_0}-\frac{ t^3}{36R_0^3}\right)\nonumber\\
& +&\hspace{-.1cm}
 \ln\left|t^2-4 R_0^2\right|\left(\frac{1}{18}-\frac{ t^4}{96R_0^4}+\frac{t^6}{2880R_0^6}\right)\,.
\end{eqnarray}
For times $t \lesssim R_0$ we find for the decoherence rate
\begin{eqnarray}\label{69}
\Gamma_\mathrm{dec}\equiv\frac{\dot{\rho}_{S,01}}{\rho_{S,01}}\approx-g^2(\phi_f-\phi_t)^2R_0^4 t\,,
\end{eqnarray}
which has to be compared with the transition frequency given by the difference between ground state and first excited 
state.
In the case of degenerate vacua, this is frequently given by the tunneling rate $\Gamma$.
We see that $\Gamma>\Gamma_\mathrm{dec}$ is in general possible for sufficiently small
times.
In cases when the nucleation radius is small or when the local minima of the
potential are very close to each other, interference effects
between different vacuum configurations are not necessarily 
negligible.
Therefore we state that there could be regions in the cosmic landscape which should be treated quantum mechanically.
The pure rate-equation approach which is frequently used is then doubtful.

For large $t\gg R_0$, the off-diagonal elements decay polynomially according to
\begin{eqnarray}\label{70}
\rho_{S,01}(t)&=&\rho_{S,01}(0)\times\\
&\times&\hspace{-.1cm} \exp\left[-\frac{4g^2}{9}(\phi_f-\phi_t)^2R_0^6\left(\frac{7}{4}+\ln\left(\frac{t}{2R_0}\right)\right)\right]\nonumber\,.
\end{eqnarray}
The result (\ref{70}) can be compared with the suppression
of interference in a well-known system, the dissipationless spin-boson model (see for example \cite{breuer}). 
Using the ohmic spectral density by $J(\omega)=\omega\exp(\omega/\Omega)$, the decoherence rate in this model is given by
\begin{eqnarray}\label{spibo}
\Gamma(t)=-\lambda\frac{1}{2}\ln(1+\Omega^2 t^2)-\lambda\ln\left(\frac{\sinh(t\pi T)}{t\pi T}\right)\,,
\end{eqnarray}
where $T$ denotes the temperature, $\lambda$ the coupling strength, and $\Omega$ the frequency cutoff.
The second term is due to the thermal contributions of the bath modes and is roughly equal to
$-\lambda t\pi T$ for $t\gg 1/T$.
Since we consider in our model only the case of vanishing temperature,
the thermal contributions vanish and we are left
with the vacuum flucutations.
Therefore we obtain only a logarithmic dependence of $\Gamma_\mathrm{dec}$
which is similar in the spin-boson model where one obtains for large times $\Gamma(t)\approx-\lambda\ln(\Omega t)$
in the limit $T\rightarrow0$.

\subsection{Localization of the growing vacuum bubble}

The reduction of the scalar field tunneling process to a two-state system is a drastic simplification.
Originally, the scalar field has infinitely many degrees of freedom.
Describing the tunneling process using the bubble radius $R$ as variable
is to a reduction to a single degree of freedom.
The two-state system considered in the preceding section, which is a further simplification of the problem,
corresponds to a superposition
of a true vacuum bubble of size 0 with a true vacuum bubble of size $R_0$.

Macroscopic objects, i.e. dust particles, are observed in well-localized states in contrast 
to microscopic particles that are often found in energy eigenstates.
The localization can be explained with the interaction of the macroscopic object
with the environment; any interference between different states
are dislocalized through continuous measurement \cite{JZKGKS03}.
A local observer has no access to the interference terms; this information
can only be obtained through an exact knowledge of the environmental
system which is in general not possible.

In the following we discuss the localization of the true vacuum
bubble where we regard the quantum mechanical variable $R$
to be continuous.
The possibility of a tunneling process back to the false vacuum will be discarded here.

In order to quantize the system given by the action (\ref{4}), we
reparametrize the action analogously to (\ref{45}) and obtain
\begin{eqnarray}\label{71}
S_R=\int d\lambda\left(\frac{4\pi R^3\epsilon}{3}\dot{t}-4\pi R^2\sigma\sqrt{\dot{t}^2-\dot{R}^2}\right)\,,
\end{eqnarray}
which is classically equivalent to
\begin{eqnarray}\label{72}
\tilde{S}_R=\int d\lambda\left[\frac{4\pi R^3\epsilon}{3}\dot{t}-2\pi R^2\sigma\left(\frac{\dot{t}^2-\dot{R}^2}{\nu}+\nu\right)\right]\,.
\end{eqnarray}
Since the kinetic term is quadratic in $\dot{R}$, we proceed with the canonical quantization procedure.
After reparametrizing (\ref{58}) and (\ref{59}) in a similar way, we obtain the following canonical momenta:
\begin{eqnarray}\label{73}
P_R&=&\frac{4\pi R^2 \sigma}{\nu}\dot{R}\,,\\
P_t&=&4\pi R^2\left(\frac{R\epsilon}{3}-\frac{\sigma\dot{t}}{\nu}\right)\\
& &-\sum_\mathbf{k}\bigg(\frac{1}{2\dot{t}^2}\dot{\psi}_\mathbf{k}\dot{\psi}_\mathbf{-k}
+\frac{1}{2}(k^2+m^2)\psi_\mathbf{k}\psi_\mathbf{-k}\nonumber
\\& &-g(\phi_t-\phi_f)\int_{|\mathbf{x}|<R} d^3xe^{i\mathbf{k x}}\psi_\mathbf{k}\bigg)\nonumber\,,\\
P_  {\psi_\mathbf{k}}&=&\frac{\dot{\psi}_\mathbf{-k}}{\dot{t}}\,.
\end{eqnarray}
The constraint Hamiltonian reads
\begin{eqnarray}\label{74}
H&=&\frac{\nu}{8\pi R^2\sigma}\Bigg\{16 \pi^2 R^4\sigma^2+P_R^2\\
& &-\bigg(P_t-\frac{4\pi R^3\epsilon}{3}+\sum_\mathbf{k}\bigg[\frac{1}{2}(k^2+m^2)\psi_\mathbf{k}\psi_\mathbf{-k}\nonumber\\
& &\hspace{-.1cm}+\frac{1}{2}
P_{\psi_\mathbf{k}}P_{\psi_\mathbf{-k}}-g(\phi_t-\phi_f)
\int_R d^3xe^{i\mathbf{k x}}\psi_\mathbf{k}\bigg]\bigg)^2\Bigg\} \approx 0\nonumber\,.
\end{eqnarray}
Since the quantization of the constraint equation (\ref{74}) does not lead to a differential equation
of Schr\"odinger type, we consider the square root of the constraint equation ignoring factor-ordering problems.
We then find
\begin{eqnarray}\label{75}
i\partial_t |\Psi\rangle&=&\bigg(\sqrt{16 \pi^2 \hat{R}^4\sigma^2+\hat{P}_R^2}-\frac{4\pi \hat{R}^3\epsilon}{3}\nonumber\\
& &+\sum_\mathbf{k}\bigg[\frac{1}{2}
\hat{P}_{\psi_\mathbf{k}}\hat{P}_{\psi_\mathbf{-k}}+\frac{1}{2}(k^2+m^2)\hat{\psi}_\mathbf{k}\hat{\psi}_\mathbf{-k}\nonumber\\
& &-g(\phi_t-\phi_f)
\int_{|\mathbf{x}|<\hat{R}} d^3xe^{i\mathbf{k x}}\psi_\mathbf{k}\bigg]\bigg)|\Psi\rangle\,,
\end{eqnarray}
where the substitution $P_t\rightarrow -i\partial/\partial_t$ was performed.
Except for the appeareance of a square root, the Hamiltonian is of standard form.
In order to simplify the problem further, we assume that the momentum $P_R$ dominates 
over the quartic term for large $R$.
This can be justified with the classical equations of motion: the radius $R$ grows 
proportionally to $t$, but $P_R$ grows proportionally to $t^3$.
With this approximation, which is valid for $t\gg R_0$, we discard all the factor ordering problems.
The system Hamiltonian simplifies to
\begin{eqnarray}\label{76}
H_0\approx |P_R|-\frac{4\pi \hat{R}^3 \epsilon}{3}\,,
\end{eqnarray}
and the corresponding Heisenberg equations of motion have the solutions
\begin{eqnarray}
\hat{R}^H(t)&=&\hat{R}_0\pm |t|\label{77}\\
\hat{P}_{R}^H(t)&=&\hat{P}_R(0)+\frac{4\pi \epsilon}{3}\left((\hat{R}_0\pm |t|)^3-\hat{R}_0^3\right)\label{78}\,.
\end{eqnarray}
Since we are interested in the localization of the growing vacuum bubbles, 
we restrict ourselves to the positive signs in equations (\ref{77}) and (\ref{78}).
The interaction now reads
\begin{eqnarray}\label{79}
H_\mathrm{int}&=&-4\pi g(\phi_t-\phi_f)\sum_\mathbf{k}\int_0^{\hat{R}}dr\frac{r}{k}\sin(kr)\psi_\mathbf{k}\nonumber\\
&\equiv&-\sum_\mathbf{k}f_k(\hat{R})\psi_\mathbf{k}\,,
\end{eqnarray}
where the radius is not fixed, in contrast to the interaction (\ref{64}).
Using equations (\ref{63}) and (\ref{66}), we find the master equation
\begin{eqnarray}\label{80}
\dot{\rho}_S&=&-i\left[H_0,\rho_S\right]\\
&+&i\int_0^t ds \sum_\mathbf{k}B_\mathrm{k}(t-s)[f_k(\hat{R}),\{f_k(\hat{R}+|t-s|),\rho_S\}]\nonumber\\
&-&\int_0^t ds\sum_\mathbf{k}A_\mathrm{k}(t-s)[f_k(\hat{R}),[f_k(\hat{R}+|t-s|),\rho_S]]\nonumber\,.
\end{eqnarray}
Since we are only interested in decoherence, we drop the unitary part as well as the terms 
describing dissipation in equation (\ref{80}).

In order to obtain an estimate for the decoherence factor, we calculate 
the $k$-dependent correlators in (\ref{80}) in the position basis, i.e.
\begin{eqnarray}\label{81}
& &\int_0^t ds\sum_\mathbf{k}A_k(t-s)f_k(R)f_k(R'+|t-s|)\nonumber\\
&=&\frac{g^2(\phi_t-\phi_f)^2}{120}\bigg\{Rt(63 R^2 R'+53{R'}^3+48 R^2 t+74{R'}^2t\nonumber\\
& &+52 R' t^2+16t^3)\nonumber\\
& &+\frac{1}{8}\left[(R-R')^2\ln(R-R')^2+(R+R')^2\ln(R+R')^2\right]\nonumber\\
& &\times(12 R^3 -14 R{R'}^2+120R R' t+80 R t^2)\nonumber\\
& &+\frac{1}{8}\left[(R-R')^2\ln(R-R')^2-(R+R')^2\ln(R+R')^2\right]\nonumber\\
& &\times(9 R^2 R'+60 R^2 t-7 {R'}^3+60{R'}^2 t +40 R't^2)\nonumber\\
& &-\frac{1}{8}\bigg[(R-R'-2t)^3\ln(R-R'-2t)^2\nonumber\\
& &+(R+R'+2t)^3\ln(R+R'+2t)^2\bigg]\times\nonumber\\
& &\times(12 R^2+7 {R'}^2+18 R't+8 t^2)\nonumber\\
& &-\frac{1}{8}\bigg[(R-R'-2t)^3\ln(R-R'-2t)^2\nonumber\\
& &-(R+R'+2t)^3\ln(R+R'+2t)^2\bigg](21 R R'+12 R t)\bigg\}\nonumber\\
&\equiv&C(R,R',t)\,.
\end{eqnarray}
Using this result it is possible to integrate the master equation (\ref{63}),
\begin{eqnarray}\label{82}
\rho(R,R',t)&=&\hspace{-.1cm}\rho(R,R',0)\exp\bigg\{-\int_0^tds\big[C(R,R,s)\\
&-&\hspace{-.1cm}C(R,R',s)-C(R',R,s)+C(R',R',s)\big]\bigg\}\,.\nonumber
\end{eqnarray}
In the limit $t\ll|R-R'|$ we find
\begin{eqnarray}\label{83}
\rho(R,R',t)&=&\rho(R,R',0)\exp\bigg[\hspace{-.1cm}-\hspace{-.05cm}\frac{g^2(\phi_t-\phi_f)^2}{8}(R\hspace{-.05cm}\hspace{-.05cm}-R')^2 t^2\nonumber\\
& &\hspace{-2.3cm}\times\left\{4(R^2\hspace{-.05cm}+\hspace{-.05cm}R R'\hspace{-.05cm}+\hspace{-.05cm}{R'}^2)\hspace{-.05cm}+\hspace{-.05cm}(R\hspace{-.05cm}+\hspace{-.05cm}R')^2\ln\frac{(R+R')^2}{(R-R')^2}\right\}\bigg]\,,
\end{eqnarray}
whereas for times $t\gg R,R'$ the non-unitary part of the density matrix is approximately given by
\begin{eqnarray}\label{84}
\rho(R,R',t)&\approx&\rho(R,R',0)\nonumber\\
& &\hspace{-2cm}\times\left|\frac{t^2(R+R'+2t)^2}{((R-R')^2-4t^2)(R+t)(R'+t)}\right|^{-\frac{1}{45}g^2(\phi_t-\phi_f)^2 t^6}\nonumber\\
& &\hspace{-2cm}\approx\rho(R,R',0)\exp\left[-\frac{g^2}{90}(\phi_t-\phi_f)^2 t^4 (R-R')^2\right]\,.
\end{eqnarray}
Compared to (\ref{70}), the suppression of the off-diagonal elements increase strongly with time,
since the vacuum bubble expands.
This is because the interaction (\ref{79}) does not assume a fixed size of the vacuum bubbles, in contrast to (\ref{64}).

\subsection{Modified tunneling rate due to external degrees of freedom}

Caldeira and Leggett showed in \cite{CL82,CL83} that the decay rate of a metastable state is modified
due to the interaction with the environment.

To obtain the modified tunneling amplitude one has to evaluate the path integral over all $\psi_\mathbf{k}$ and normalize
the resulting expression such that the impact of the interaction vanishes if $g$ tends to zero.
Since the interaction (\ref{59}) is bilinear, this will reduce to a ratio of two functional integrals.

It is important to include a renormalization term,
since the environment causes a frequency shift of the potential.
After switching to imaginary time $T=-it$ we find the euclidean action
\begin{eqnarray}\label{85}
S_\mathrm{E}=S_\mathrm{E,R}+S_\mathrm{E,bath}+S_\mathrm{E,int}+S_\mathrm{E,ren}
\end{eqnarray}
with
\begin{eqnarray}\label{86}
S_\mathrm{E,R}&=&-\int_0^{T_0}dT\left(\frac{4\pi\epsilon}{3}R^3-4\pi\sigma R^2\sqrt{1+\dot{R}^2}\right)\,,\nonumber\\
S_\mathrm{E,bath}&=&\hspace{-.1cm}\int_0^{T_0}dT\sum_\mathbf{k}\frac{\mathcal{V}}{2}\left(\dot{\psi}_\mathbf{k}\dot{\psi}_\mathbf{-k}+
(k^2+m^2)\psi_\mathbf{k}\psi_\mathbf{-k}\right)\,,\nonumber\\
S_\mathrm{E,int}&=&-\int_0^{T_0}dT\sum_\mathbf{k}f_k(R)\psi_\mathbf{k}\nonumber\,,\\
S_\mathrm{E,ren}&=&\int_0^{T_0}dT\sum_\mathbf{k}\frac{f_k^2(R)}{2 \mathcal{V}(k^2+m^2)}\,,
\end{eqnarray}
where $f(R)$ was defined in equation (\ref{79}) and $\mathcal{V}$ denotes again the quantization volume of the environmental modes.
The euclidean time at which the bubble radius vanishes is denoted with $T_0$ and coincides with $R_0$ in the limit
of vanishing system-environment interaction.
The ratio of the functional integrals can be evaluated excactly since the action is quadratic in the $\psi_\mathbf{k}$'s (see \cite{FEYN65}).
We find
\begin{eqnarray}\label{87}
\frac{\int \Pi_\mathbf{k}\mathcal{D}{\psi_\mathbf{k}}\exp(-S_\mathrm{E})}
{\int \Pi_\mathbf{k}\mathcal{D}{\psi_\mathbf{k}}\exp(-S_\mathrm{E}(g=0))}=\exp(-S_\mathrm{eff})
\end{eqnarray}
with
\begin{eqnarray}\label{88}
S_\mathrm{eff}&=&S_\mathrm{E,R}+S_\mathrm{E,ren}\nonumber\\
& &-\int_0^{T_0}dT\int_0^{T_0}dT'\sum_\mathbf{k}\frac{f(R)f(R')}
{4\mathcal{V}\sqrt{k^2+m^2}}\nonumber\\
& &\times\frac{\cosh\left(\sqrt{k^2+m^2}\left(|T-T'|-\frac{T_0}{2}\right)\right)}{\sinh\left(\sqrt{k^2+m^2}\frac{T_0}{2}\right)}\,.
\end{eqnarray}
Varying the effective action, we find the equations of motion to be
\begin{eqnarray}\label{91}
\frac{d}{dT}\left(\frac{4\pi\sigma R^2\dot{R}}{\sqrt{1+\dot{R}^2}}\right)&=&-4\pi\epsilon R^2
+8\pi\sigma R\sqrt{1+\dot{R}^2}\nonumber\\
& &\hspace{-3cm}+\sum_\mathbf{k}\frac{f_k(R)\partial_R f_k(R)}{\mathcal{V}(k^2+m^2)}-\int_0^{T_0}
 dT'\sum_\mathbf{k} \frac{f_k(R')\partial_{R}f_k(R)}{2\mathcal{V}\sqrt{k^2+m^2}}\nonumber\\
 & &\hspace{-3cm}\times\frac{\cosh\left(\sqrt{k^2+m^2}\left(|T-T'|-\frac{T_0}{2}\right)\right)}{\sinh\left(\sqrt{k^2+m^2}\frac{T_0}{2}\right)}\,.
\end{eqnarray}
\begin{figure}
\includegraphics{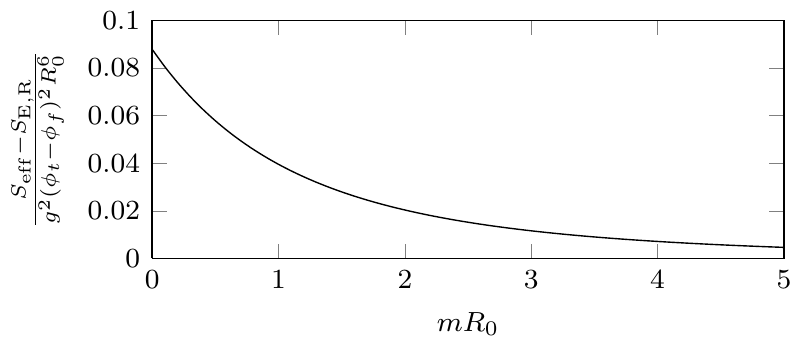}
\caption{The system-enviroment interaction decreases with growing mass of the environmental field.
}\label{fig6}
\end{figure}
The terms involving the interaction can be interpreted as nonlocal friction terms leading
to a reduction of the tunneling rate \cite{CL82}.
Although in general one would have to solve equation (\ref{91}) in order to find
the numerical value of the instanton action, we neglect here the back reaction
of the environment on the bubble and set $T_0=R_0$.
Substituting
\begin{eqnarray}\label{89}
\sum_\mathbf{k}\rightarrow \frac{\mathcal{V}}{(2\pi)^3}\int d^3k
\end{eqnarray}
and evaluating the integals numerically, we find for $m=0$ the correction to the imaginary part of the euclidean classical action (\ref{4})
to be
\begin{eqnarray}\label{90}
S_\mathrm{eff}-S_\mathrm{E,R}\approx0.088 g^2(\phi_t-\phi_f)^2R_0^6\,.
\end{eqnarray}
The correction for arbitrary $m$ is plotted in Fig. (\ref{fig6}).
Since the nonlocal terms as well as the renormalization term in the action (\ref{88}) decrease with
growing $m$, we find that the suppression of the tunneling process for large masses is weaker than it is for 
small masses.

\section{conclusions}

We have discussed in various settings the behaviour of vacuum tunneling in curved backgrounds.
In particular the solutions concerning de Sitter space should be important for the 
investigation of the cosmic landscape.

Futhermore we have argued that the instanton picture might not be applicable
in situations where decoherence is weak.
This is the case for sufficiently small nucleation radii or small system-environment 
couplings.
We have chosen the interaction between system and environment to be linear in the environmental degrees of freedom,
an assumption which has been applied to varius models that can be described by a macroscopic variable \cite{CL83}.
The motivation behind this linear coupling is, that every single environmental degree of
freedom is only weakly perturbed by the system.
This does not mean that the effect on the system is weak, since infinitely many degrees of freedom are involved.

The specific form how the macroscopic variable enters the interaction was derived
from a generic bilinear and locally Lorentz-invariant interaction between system field and environmental
field.
In contrast to the treatment of decoherence in quantum mechanical models, we did not need to
assume a particular form of the spectral density, i.e. $J(\omega)\propto \omega^s$.

An important aspect for future research addressing the tunneling in curved
backgrounds could be the inclusion of backreaction due to gravity.
In this case, the background will depend nontrivially on the surface tension
and the difference between the energy densities in the local minima.

\section*{ACKNOWLEDGEMENTS}

I want to thank Claus Kiefer, Julian Adamek, Rochus Klesse and Ralf Sch\"utzhold for
many helpful discussions.
This work was supported by the Bonn-Cologne Graduate School.

\bibliographystyle{plain}

\end{document}